\documentclass[preprint,12pt,a4paper]{elsarticle}

\usepackage{graphicx}
\usepackage{tikz}
\usetikzlibrary{positioning}
\usepackage{float}
\usepackage{listings}
\usepackage{graphicx}
\usepackage[font=small]{caption}
\usepackage{placeins}   
\usepackage[a4paper, margin=1in]{geometry}
\usepackage{lineno,hyperref}
\modulolinenumbers[5]
\usepackage{xcolor}
  
\usepackage{multirow}
\usepackage[numbers]{natbib}

\begin{document}

\begin{frontmatter}

\title{"Digital Camouflage": The LLVM Challenge in LLM-Based Malware Detection}

\author{
  Ekin Böke \\ 
  {\small Technische Universität Berlin, 10623 Berlin, Germany} \\
  {\small Applied Machine Learning Department / Fraunhofer Heinrich-Hertz-Institute, 10587 Berlin, Germany}
  \vspace{2em} \\  
  Simon Torka \\ 
  {\small Distributed Artificial Intelligence Laboratory / Technische Universität Berlin , 10623 Berlin Germnay}
}

\begin{abstract}

Large Language Models (LLMs) have emerged as promising tools for malware detection by analyzing code semantics, identifying vulnerabilities, and adapting to evolving threats. However, their reliability under adversarial compiler-level obfuscation is yet to be discovered.
In this study, we empirically evaluate the robustness of three state-of-the-art LLMs: ChatGPT-4o, Gemini Flash 2.5, and Claude Sonnet 4 against compiler-level obfuscation techniques implemented via the LLVM infrastructure. These include control flow flattening, bogus control flow injection, instruction substitution, and split basic blocks, which are widely used to evade detection while preserving malicious behavior. We perform a structured evaluation on 40~C functions (20 vulnerable, 20 secure) sourced from the Devign dataset and obfuscated using LLVM passes. Our results show that these models often fail to correctly classify obfuscated code, with precision, recall, and F1-score dropping significantly after transformation. This reveals a critical limitation: LLMs, despite their language understanding capabilities, can be easily misled by compiler-based obfuscation strategies. To promote reproducibility, we release all evaluation scripts, prompts, and obfuscated code samples in a public repository. We also discuss the implications of these findings for adversarial threat modeling, and outline future directions such as software watermarking, compiler-aware defenses, and obfuscation-resilient model design.

\end{abstract}


\end{frontmatter}

\pagenumbering{arabic}

\newpage

\tableofcontents

\newpage
\pagenumbering{arabic} 

\section{Introduction}
\label{Introduction}

Malware detection and mitigation are important parts of modern cybersecurity efforts. During the past decade, LLMs have emerged as powerful tools in this field. These models excel at identifying deviations from the norm, referred to as anomalies. Recent research even shows the possibility of using LLMs as code executers and analyzing code and detecting anomalies by executing code and comparing expected behavior and actual output~\cite{lyu2024codeexecutors}.

Unlike traditional signature-based detection, LLMs leverage their training on diverse datasets to identify threats, even in previously unseen contexts. This makes them particularly useful in dynamic and rapidly evolving threat landscapes. \textcolor{blue}{Yet, their effectiveness crucially depends on the structure and transparency of the code they analyze. A key factor is the role of compiler-generated intermediate representations, which form the level where both optimizations and obfuscation techniques are applied.}

\textcolor{blue}{LLVM, a widely used compiler framework, first translates C source code (\texttt{.c}) into an intermediate representation known as LLVM Intermediate Representation (LLVM IR)~\cite{llvm}. The LLVM IR exists in two equivalent forms: as a human-readable, assembler-like format (\texttt{.ll}) and as a compact, binary-encoded variant (\texttt{.bc}).}
\textcolor{blue}{Optimization and obfuscation passes act directly at this level before the program is translated into platform-specific binary files such as object files (\texttt{.o}) or executable files (\texttt{.exe}).}
\textcolor{blue}{LLVM IR encodes program semantics at a low but still abstract level, making it suitable not only for standard compiler transformations but also for obfuscation techniques such as control flow smoothing, command substitution, or the insertion of dead code, which can significantly obscure syntactic and semantic features. This makes it difficult for LLM-based detection models to reliably identify malicious code in obfuscated LLVM code, which is a key weakness of current approaches.}

This growing use of sophisticated obfuscation calls for a deeper examination of how current malware detection techniques especially those based on LLMs cope with adversarial transformations.

To establish this context, Chapter~\ref{EvolutionofLLMbasedMalwareDetection} provides a background on malware detection techniques and their evolution over time, highlighting how advances in natural language processing (NLP) enabled the emergence of LLM-based detection methods.

Chapter~\ref{MalwareObfuscationTechniques} introduces malware obfuscation techniques and evasion strategies commonly used to bypass detection systems. A key focus is placed on the role of LLVM in code transformation, particularly how its optimization and obfuscation capabilities can be leveraged to alter malicious code while maintaining functionality.

\textcolor{blue}{Building on this foundation, Chapter~\ref{AdversarialThreatModel} defines the adversarial threat model underpinning our study. We consider a Man-At-The-End (MATE) attacker. In this scenario, the defender controls the analysis environment: both the compiled binaries and their corresponding LLVM Intermediate Representation (LLVM IR). The attacker, on the other hand, has direct access to the software and deliberately employs compiler-level obfuscation techniques to evade both static and dynamic scrutiny.}

\textcolor{blue}{This scenario reflects realistic operational conditions. In real-world investigations, security analysts often encounter obfuscated binaries obtained from malware repositories, phishing campaigns, or compromised systems. However, these raw binary files cannot be directly analyzed using LLMs, which are specialized for natural language processing and cannot interpret binary code without specialized training. In contrast, the LLVM IR format, especially its textual \texttt{.ll} representation offers a middle ground: it encodes low-level semantics in a format that is both human-readable and machine-ingestable by LLMs.}

\textcolor{blue}{While \texttt{.ll} files are typically not distributed with binary payloads, they can be recovered through leaked source code, builder toolkits, or reverse engineering. For example, Toor~\cite{toor2022ghidra} demonstrates that tools like Ghidra~\cite{ghidra} specifically Ghidrall can reconstruct LLVM IR from binaries using Ghidra’s P-Code and decompilation trees, enabling downstream analysis even in the absence of original source code.}

\textcolor{blue}{To systematically evaluate the impact of such transformations on LLM-based vulnerability detection, we design a controlled benchmarking setup leveraging representative compiler-level obfuscation passes and real-world vulnerable code samples.}

\begin{figure}[!htbp]
  \centering
  \includegraphics[width=\textwidth]{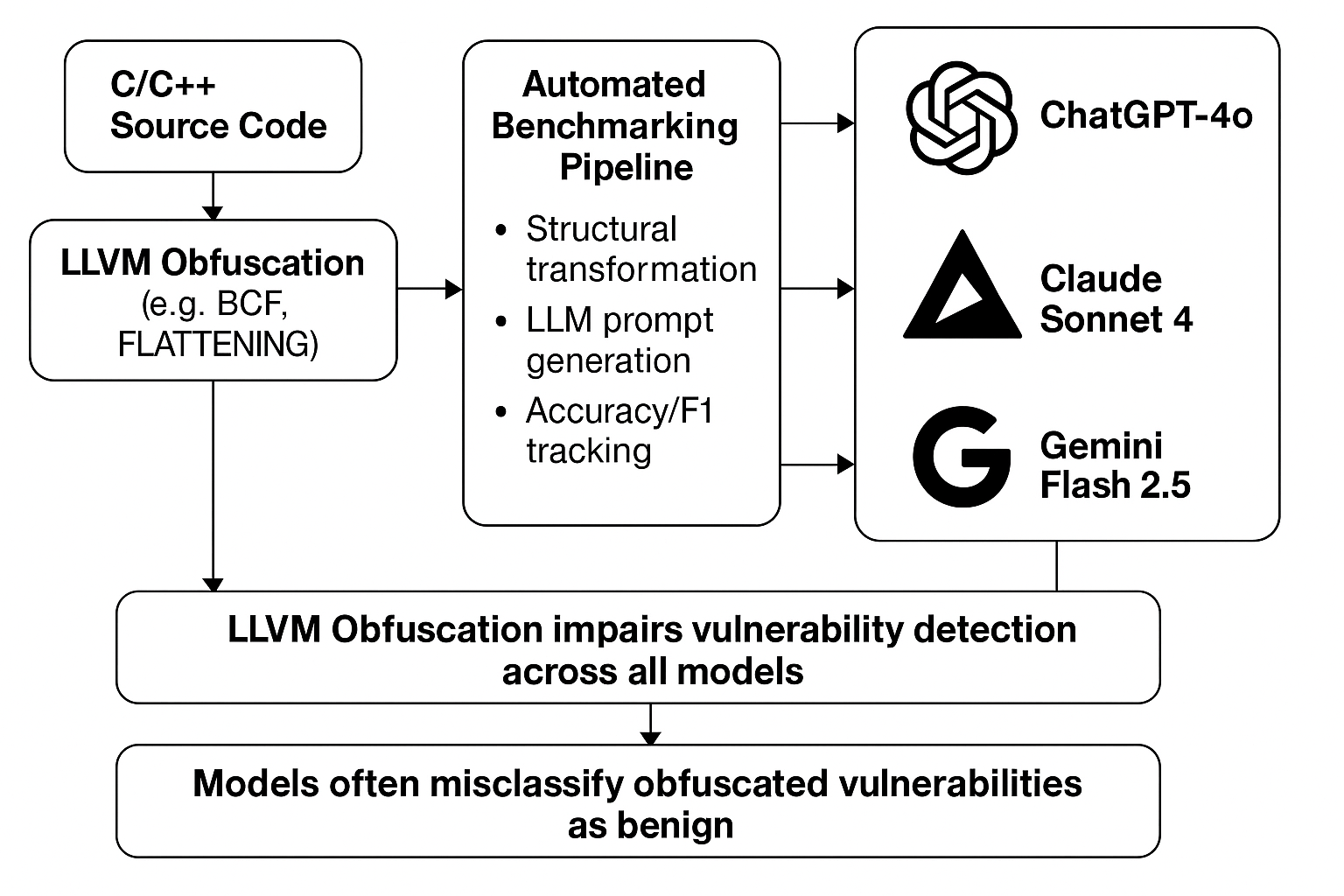} 
  \caption{Overview of our benchmarking pipeline.}
  \label{fig:pipeline_overview}
  
  \vspace{1mm}
  {\small \textit{Logos are trademarks of their respective owners. Used here for illustrative purposes only.}}
\end{figure}

\textcolor{blue}{Chapter~\ref{LLVM-basedObfuscationandBenchmarking} presents a comprehensive benchmarking study that evaluates the robustness of three state-of-the-art LLMs: \textit{ChatGPT-4o}~\cite{openai_chatgpt4o}, \textit{Gemini Flash 2.5}~\cite{gemini2025}, and \textit{Claude Sonnet 4}~\cite{claude2025} against a set of adversarially obfuscated code samples. To clarify our experimental setup, Figure~\ref{fig:pipeline_overview} illustrates the full pipeline from code selection to evaluation.}

\FloatBarrier

\textcolor{blue}{In our obfuscation pipeline, we rely on the main branch of the publicly available \textit{Obfuscator-LLVM} project~\cite{obfuscatorllvm2014}, compatible with LLVM~14. This version supports four stable and well-documented transformation passes: control-flow flattening, bogus control-flow insertion, instruction substitution, and basic-block splitting. These passes are widely studied and have been observed in both academic literature and real-world malware~\cite{revay2023cffvb}, offering a representative sample of contemporary obfuscation strategies.}

\textcolor{blue}{To provide meaningful input to the models, we sampled 40 functions (20 labeled as vulnerable, 20 as safe) from the Devign dataset~\cite{zhou2019devign}. Minimal modifications were made to ensure compilability, such as inserting missing type declarations, without changing the original program logic or semantics. A detailed rationale for this selection, along with implementation details, is provided in Section~\ref{ObfuscationPipeline}.}

\textcolor{blue}{Each function, both in its original and obfuscated form, was evaluated in a zero-shot setting using the three LLMs. Since LLMs are inherently generative and non-deterministic across sessions, we repeated each evaluation three times using separate chat sessions to ensure result consistency.}

\textcolor{blue}{We evaluate model outputs using standard classification metrics: precision, recall, and F1 score  before and after obfuscation. The results reveal a substantial degradation in performance for certain models, underscoring the vulnerability of LLM-based detection systems to compiler-level adversarial transformations.}

\textcolor{blue}{Chapter~\ref{Limitations} discusses the limitations of our study, including dataset size, pass diversity, and the use of zero-shot prompting. These considerations contextualize the scope and generalizability of our findings.}

\textcolor{blue}{Finally, Chapter~\ref{ConclusionOutlook} summarizes our conclusions and highlights promising directions for future work. These include the use of compiler-level obfuscation for software protection, software watermarking, adversarial hardening of LLM-based security tools, and training models directly on low-level representations such as LLVM IR to improve resilience against transformation-based attacks.}

\section{Evolution of LLM based Malware Detection}
\label{EvolutionofLLMbasedMalwareDetection}

Malware detection has been a fundamental aspect of cybersecurity since the widespread adoption of the Internet, growing alongside the increasing complexity of malicious software. Traditional approaches relied on simpler signature-based methods. Although effective against previously identified threats, this approach faced significant challenges with zero-day attacks, where no prior signature exists, and polymorphic malware, which alters its code structure while maintaining its functionality. A key limitation of signature-based methods is their inherently reactive nature as malware techniques became increasingly complex, the field has found the solution in machine learning (ML). ML began to address the limitations of traditional methods by learning patterns in data that can be used for classification purposes. Common techniques included static analysis, which involves analyzing code structure and features without executing the program; dynamic analysis, which observes runtime behavior to identify malicious actions; and hybrid analysis, which combines both methods. Early machine learning models, such as Support Vector Machines (SVMs) and Random Forests, proved effective in classification with extracted features from code. However, these models struggled to generalize to large datasets. The rise of big data brought with it challenges, as highlighted by studies such as~\cite{feature}, which underlined the importance of robust feature engineering. The study by Raff et al.~\cite{raff2017malware} introduced MalConv, a deep learning model designed for malware detection by analyzing raw binary files. Although not a LLM, MalConv demonstrated the effectiveness of direct input analysis, bypassing traditional manual feature extraction. This approach shares conceptual similarities with models based on NLP, which process long sequences and capture complex dependencies. Recent advances in LLMs build on these ideas, leveraging their ability to analyze entire code structures holistically, identify malicious intent, and capture global patterns. For example, the study by Singh et al.~\cite{singh2024malicious} demonstrates the application of LLMs in the analysis of Java code for malicious behaviors, highlighting their capacity to handle complex code patterns. By integrating improvements in sequence modeling and contextual understanding, LLMs have opened new possibilities for malware analysis, offering enhanced capabilities to detect obfuscated or complex malicious behavior.

\subsection{LLM-Based Malware Detection}
\label{LLM-BasedMalwareDetection}

LLMs, such as GPT-3~\cite{brown2020language} and BERT~\cite{devlin2018bert},  have influenced malware detection using their ability to process large-scale statistical representations of source code. Although LLMs do not inherently "understand" code as humans do, they can capture patterns and relationships within code syntax and semantics through pretraining on extensive datasets.
\textcolor{blue}{Transformer-based models such as CodeBERT~\cite{Feng:codebert} and CodeGPT~\cite{lu2021codexglue} have been fine-tuned on code corpora to support downstream tasks like vulnerability detection and automated code repair, though they are typically used in supervised or pipeline-based settings rather than through direct user interaction.}

A key strength of LLMs lies in processing code as token sequences, similar to NLP tasks. Pre-training on large data sets of source code and programming documentation enables these models to capture latent patterns and understand software behaviors. LLMs lack explicit program flow modeling and semantic understanding at the level of tools such as abstract syntax trees (ASTs) or control flow graphs (CFGs). Operating primarily at the token or sequence level, they can overlook deeper structural relationships unless fine-tuned on task-specific datasets designed to emphasize these aspects.

One such domain-specific model, CodeBERT~\cite{Feng:codebert}, exemplifies how LLMs tailored to programming languages can overcome some of these limitations. By leveraging token embeddings and attention mechanisms, CodeBERT effectively detects patterns in code, including obfuscated or incomplete samples. While it does not explicitly model program flow semantics, its capacity to capture token relationships has demonstrated significant advantages in tasks like vulnerability detection and code understanding. For example, in the work by Li et al.~\cite{li2024automated}, CodeBERT was successfully applied to detect SQL injection vulnerabilities. By embedding code representations and utilizing datasets from real-world repositories, the model achieved notable improvements in detection performance and training efficiency, outperforming traditional static application security testing (SAST) tools. This reinforces the potential of deep learning models like CodeBERT to transform security analysis and elevate software development practices.

Building on the success of CodeBERT, MalBERT~\cite{Rahali2021MalBERT} is a domain-specific adaptation designed for malware detection by leveraging its ability to analyze token relationships in code. The model is fine-tuned using the Androzoo public dataset~\cite{Allix2016Androzoo} to capture patterns and signatures characteristic of malicious code in Android applications, such as API misuse, control flow obfuscation, and embedded malicious payloads. This domain-specific pre-training enables MalBERT to effectively identify malware, including detecting suspicious permissions or obfuscated strings. Its success demonstrates how domain-specific adaptations of LLMs can enhance their effectiveness in niche areas, addressing limitations seen in general-purpose models.

Similarly, in a study by Demirci et al.~\cite{9785789}, a domain-specific GPT-2 model (DSLM-GPT2) was pre-trained on assembly instructions to capture contextual dependencies. Although DSLM-GPT2 achieved an F1 score of 86.0\% in sentence-level analysis, it was outperformed by a document-level analysis model (DLAM) based on stacked Bidirectional Long Short-Term Memory (BiLSTM) networks, which achieved an F1 score of 98.3\%. This highlights the importance of holistic understanding of the data and the ability to model sequential dependencies effectively. The findings underscore that, while LLMs like GPT-2 excel at capturing token-level relationships, complementary approaches like BiLSTM, which explicitly model sequential dependencies, may better address tasks requiring a broader contextual understanding, such as static analysis of code.

LLMs show great potential in cybersecurity due to their ability to generalize across various programming languages and recognize patterns resembling novel malware. Studies such as~\cite{AkinsowonJiang2024} have demonstrated that LLMs can differentiate known malware from novel families, thus achieving high accuracy rates even with previously unseen malware. 

Similarly, research like~\cite{li2022multiview} highlights that models can detect adversarial malware variants across multiple subtypes, highlighting their ability to generalize and identify new threats. In the paper Adversarially Robust Multi view Malware Defense (ARMD) a new multi view learning framework that enhances the robustness of deep learning based malware detectors against adversarial variants is presented. By using the multiple views of a malware file, such as the source code view alongside the binary view, ARMD exploits the fact that adversarial modifications often leave certain characteristics in the source code unchanged. This approach significantly improves the detection of adversarial malware variants. Experiments on three renowned open source deep learning-based malware detectors across six common malware categories demonstrate that ARMD can improve adversarial robustness by up to seven times. This further solidifies the idea that the effectiveness of LLMs is highly dependent on the diversity and completeness of their training data. While they excel at recognizing familiar patterns, their ability to detect entirely new or significantly different malware without additional fine-tuning may be limited in the case of data sparcity. Challenges such as code obfuscation and the substantial computational resources required for real-time deployment further hinder their widespread use.

\textcolor{blue}{While such hybrid frameworks that integrate LLMs with holistic analysis techniques offer  improvements in feature representation and generalization, they often fall short when facing obfuscated malware. In particular, LLVM-based code transformations can severely distort syntactic and semantic patterns that LLMs rely on. These challenges underscore the need to systematically examine how resilient current LLM-based approaches are when confronted with obfuscation.}

\textcolor{blue}{To this end, we deliberately chose to evaluate general-purpose LLMs such as ChatGPT-4o~\cite{openai_chatgpt4o}, Gemini Flash 2.5~\cite{gemini2025}, and Claude Sonnet 4~\cite{claude2025} rather than specialized models such as CodeBERT~\cite{Feng:codebert} or MalBERT~\cite{Rahali2021MalBERT}. While the latter have been fine-tuned for specific downstream tasks such as code search, summarization, or vulnerability detection, the generalist models we investigate represent the current state of the art in universal, dialog-oriented LLMs. These models are increasingly applied in real-world software engineering practice for tasks ranging from code review and debugging to security analysis. Consequently, their robustness against code obfuscation is of immediate relevance to practitioners and the broader research community. To the best of our knowledge, this is the first systematic study assessing how LLVM-based obfuscation impacts malware detection with general-purpose LLMs, thereby filling a gap in existing research that has so far focused primarily on unobfuscated code or on specialized, task-specific models.}

\clearpage

\section{Malware Obfuscation Techniques}
\label{MalwareObfuscationTechniques}

Malware obfuscation techniques are used by attackers to evade detection and analysis. By modifying the structure, appearance, and behavior of malicious code, obfuscation can make it significantly more difficult for security analysts to recognize and eliminate threats. These techniques range from simple syntactic changes to advanced transformations that adapt according to the execution environment. Understanding them is essential for developing effective countermeasures and improving malware detection strategies. In this chapter, we will explore various malware obfuscation techniques. Detailed examples of obfuscation techniques are provided in ~\ref{CodeObfuscationExamples}.

\subsection{Introduction to Malware Obfuscation}
\label{IntroductiontoMalwareObfuscation}

Malware obfuscation is a set of techniques designed to conceal malicious code and evade detection. These techniques modify the structure, appearance, or behavior of malware without altering its functionality, complicating both static and dynamic analysis. 
Obfuscation methods can be broadly categorized into \textbf{syntactic}, \textbf{semantic}, and \textbf{behavioral} transformations:

\begin{itemize}
    \item \textbf{Syntactic Obfuscation}: Alters the structure of code while keeping its logic intact. This includes \textit{variable renaming}, \textit{dead code insertion}, and \textit{formatting changes} to mislead static analysis tools. Although these transformations do not modify the program’s behavior, they can significantly hinder readability and make reverse engineering more time-consuming.
    
    \item \textbf{Semantic Obfuscation}: Modifies how code executes by introducing \textit{control flow distortions}, \textit{opaque predicates}, and \textit{instruction substitution}, making analysis significantly harder for reverse engineering and decompilation tools. These techniques not only obscure the program’s logic but also increase the computational overhead required for security tools to analyze the code effectively.

    \item \textbf{Behavioral Obfuscation}: Affects runtime behavior through \textit{encryption}, \textit{packing}, and \textit{polymorphism/metamorphism}, ensuring that the malicious payload remains hidden until the time of execution. By dynamically decrypting or unpacking code in memory, these techniques evade static analysis and signature-based detection. This approach allows malware to adapt and mutate, making it a challenge for traditional security tools to recognize and neutralize threats. Additionally, some advanced obfuscation methods incorporate environment-aware execution, ensuring that the malicious payload activates only under specific conditions, further complicating detection efforts.

\end{itemize}

These techniques force security tools to rely on more advanced heuristics, AI-driven anomaly detection, or runtime monitoring to catch malicious activity. 

\subsection{LLVM and Its Role in Code Transformation}
\label{LLVMandItsRoleinCodeTransformation}

LLVM provides a powerful compiler infrastructure for creating a variety of compiler tools, such as optimizers, debuggers, and static analyzers. Due to its modular architecture, LLVM has become a widely used platform for building sophisticated software tools, including those designed for obfuscating and transforming code. The flexibility of LLVM makes it particularly appealing for tasks requiring fine-grained control over code manipulation~\cite{lattner2004llvm}. 

LLVM operates using an Intermediate Representation (IR), a platform-independent abstraction layer that decouples source code from target machine code. This design enables cross-language compatibility and allows for advanced optimizations across multiple programming languages. LLVM IR is a low-level, strongly-typed representation based on Static Single Assignment (SSA) form, ensuring efficient transformations and analysis while serving as a bridge between high-level code and machine instructions.

To better understand how LLVM IR works, we consider a simple C program, which is shown in Table~\ref{tab:c_code}. 

\begin{table}[H]
    \centering
    \begin{tabular}{|l|}
        \hline
        \texttt{\#include <stdio.h>} \\ \hline
        \texttt{int calculate(int a, int b) \{} \\ 
        \texttt{\quad return a + b;} \\ 
        \texttt{\}} \\ \hline
        \texttt{int main() \{} \\ 
        \texttt{\quad int result = calculate(5, 3);} \\ 
        \texttt{\quad printf("Result: \%d\textbackslash n", result);} \\ 
        \texttt{\quad return 0;} \\ 
        \texttt{\}} \\ \hline
    \end{tabular}
    \caption{Simple C code}
    \label{tab:c_code}
\end{table}

The corresponding LLVM IR representation, obtained by compiling the code with the \texttt{-emit-llvm} flag in Clang, is provided in Table~\ref{tab:llvm_ir}.

\begin{table}[h]
    \centering
    \begin{tabular}{|l|}
        \hline
        \texttt{; Function to add two integers} \\ \hline
        \texttt{define i32 @calculate(i32 \%a, i32 \%b) \{} \\ 
        \texttt{\quad entry:} \\ 
        \texttt{\quad \%sum = add i32 \%a, \%b} \\ 
        \texttt{\quad ret i32 \%sum} \\ 
        \texttt{\}} \\ \hline
        \texttt{; Main function} \\ \hline
        \texttt{define i32 @main() \{} \\ 
        \texttt{\quad entry:} \\ 
        \texttt{\quad \%result = call i32 @calculate(i32 5, i32 3)} \\ 
        \texttt{\quad \%formatStr = alloca [10 x i8], align 1} \\ 
        \texttt{\quad store [10 x i8] c"Result: \%d\textbackslash0A\textbackslash00", ptr \%formatStr} \\ 
        \texttt{\quad call i32 (ptr, ...) @printf(ptr \%formatStr, i32 \%result)} \\ 
        \texttt{\quad ret i32 0} \\ 
        \texttt{\}} \\ \hline
        \texttt{declare i32 @printf(ptr, ...)} \\ \hline
    \end{tabular}
    \caption{LLVM IR Representation of a Simple C code}
    \label{tab:llvm_ir}
\end{table}

LLVM IR incorporates several key features that make it a powerful intermediate representation for compiler optimizations and analysis. One of its core principles is SSA form, where each variable is assigned exactly once. This ensures better optimization and analysis, as seen in the instruction \%sum = add i32 \%a, \%b, where \%sum remains immutable throughout execution. LLVM IR is also strongly typed, meaning that each variable has a specific type. For example, in the function @add, the parameters \%a and \%b are explicitly declared as i32 (32-bit integers), enforcing strict type safety. Additionally, LLVM IR employs explicit memory management, avoiding high-level memory abstractions. Instead, it uses the alloca instruction for manual memory allocation, as demonstrated in \%formatStr = alloca [10 x i8], align 1, which reserves stack space for a string. Function calls in LLVM IR are also explicitly defined; for instance, the instruction \%result = call i32 @calculate(i32 5, i32 3) clearly invokes the @calculate function, providing a low-level representation of function execution. Finally, LLVM IR supports instruction-level optimizations, including function inlining, constant propagation, and dead code elimination, ensuring efficient machine code generation before execution. These features make LLVM IR a robust and flexible foundation for program transformations, security mechanisms, and advanced compiler optimizations. 

The ability to manipulate this intermediate representation allows LLVM to serve as the foundation for complex code transformations, enabling malware authors to employ advanced obfuscation techniques to evade detection by static analysis tools, as in the work of Sharif et al.~\cite{sharif2008impeding}. They introduce a technique called Conditional Code Obfuscation (CCO), which alters the execution of a program according to its environment.
Malware authors can leverage Conditional Code Obfuscation (CCO) to alter a program's behavior based on its execution environment, allowing malicious code to evade detection in sandboxes or debuggers while operating normally in real-world scenarios. The authors implemented CCO using an LLVM-based framework, enabling obfuscation techniques to be applied directly at the compiler level. Their evaluation against common malware analysis tools, including IDA Pro~\cite{idapro} and static disassemblers, demonstrated that these obfuscation methods significantly increase analysis time and obstruct the reconstruction of control flow graphs, which static analysis tools rely on. The paper concludes that CCO and similar obfuscation techniques present a formidable challenge to modern malware analysis. While CCO, focuses on dynamically altering program execution, another prominent obfuscation approach at the compiler level is Obfuscator-LLVM (O-LLVM)~\cite{junod2015obfuscator}. Developed as an extension to the LLVM compilation framework, O-LLVM systematically applies obfuscation transformations at the IR level, integrating them into the compilation process rather than modifying the binary afterwards.

Unlike CCO, which relies on runtime environment checks to alter execution behavior, O-LLVM  applies static code transformations that increase the difficulty of reverse engineering. By leveraging LLVM’s modular compilation pipeline, O-LLVM implements obfuscation techniques designed to hinder analysis tools such as IDA Pro and Ghidra~\cite{ghidra}. These include:

\begin{itemize}

    \item \textbf{Control Flow Flattening:} Restructures a program’s control flow into a dispatcher-based model, making it difficult to reconstruct the original branching structure.
    
    \item \textbf{Instruction Substitution:} Replaces standard operations with functionally equivalent but more complex alternatives, disrupting pattern-based analysis.
    
    \item \textbf{Bogus Control-Flow Insertion:} Introduces opaque predicates to create misleading execution paths that confuse decompilers.
    
    \item \textbf{Procedure Merging:} Combines multiple functions into a single unit, obscuring logical program structure.
    
    \item \textbf{Tamper-Proofing Mechanisms:} Embeds runtime integrity checks to detect unauthorized modifications and potential analysis attempts.
    
\end{itemize}

Junod et al.~\cite{junod2015obfuscator} evaluate the effectiveness of these transformations and highlight their ability to significantly increase reverse engineering effort while introducing only moderate runtime overhead. Their findings demonstrate that compiler-assisted obfuscation provides a structured alternative to traditional binary obfuscation techniques by ensuring that transformations are deeply integrated into the program’s structure before code generation.

As LLVM-based obfuscation techniques continue to evolve, they play a crucial role in the broader field of software security by complicating static analysis and reverse engineering. The continued development of these methods highlights the ongoing arms race between obfuscation and de-obfuscation, reinforcing the need for more advanced reverse engineering tools in modern security research. \textcolor{blue}{
The arms race is evident not only in research but also in the threat landscape, where attacker use similar techniques to complicate analysis and detection (see Chapter~\ref{AdversarialThreatModel}).
}

\section{Adversarial Threat Model}
\label{AdversarialThreatModel}

\textcolor{blue}{This study is grounded in a realistic MATE threat model, inspired by the tactics of APTs. We assume the attacker has full control over the software build environment, including the ability to alter the source code and apply compiler-level obfuscation during compilation. Such an assumption is motivated, as prior research shows that modern malware and mutation engines routinely employ such transformations. Brezinski's metamorphic malware survey highlights that malware authors widely use techniques such as bogus control flow insertion, instruction substitution, and basic block transformations to avoid detection and complicate analysis pipelines~\cite{brezinski2023metamorphic}. Complementing this perspective, Zhang et al. demonstrate in their Khaos framework that LLVM-based obfuscations degrade the effectiveness of binary diffing techniques, a core method for vulnerability discovery~\cite{zhang2023khaos}.}

\textcolor{blue}{Evidence from the field further substantiates this threat model: the \textit{ANEL} backdoor, attributed to APT10, has been shown to employ compiler-level obfuscations including opaque predicates and control-flow flattening to resist both static and dynamic analysis~\cite{vmware2019apt10}. Taken together, these findings indicate that our assumed adversarial capabilities are not only realistic but also reflective of techniques already observed in academic studies and real-world threat campaigns.}

\textcolor{blue}{By aligning our threat model with these operational techniques, we ensure that the obfuscated code evaluated in our pipeline closely reflects the complexity and evasiveness of real-world adversaries. In turn, this allows us to rigorously assess whether LLMs can reliably identify vulnerabilities \textit{under adversarial compilation conditions}, a crucial benchmark for their deployment in security-critical domains.}

\section{LLVM-based Obfuscation and Benchmarking}
\label{LLVM-basedObfuscationandBenchmarking}

Code obfuscation plays a critical role in cybersecurity, serving both defensive and offensive purposes. While developers may use obfuscation to protect intellectual property and improve software resilience, adversaries often exploit the same techniques to evade AI-based analysis. With the growing adoption of LLMs in automated vulnerability detection, understanding how obfuscation interferes with their capabilities is an essential area of research.

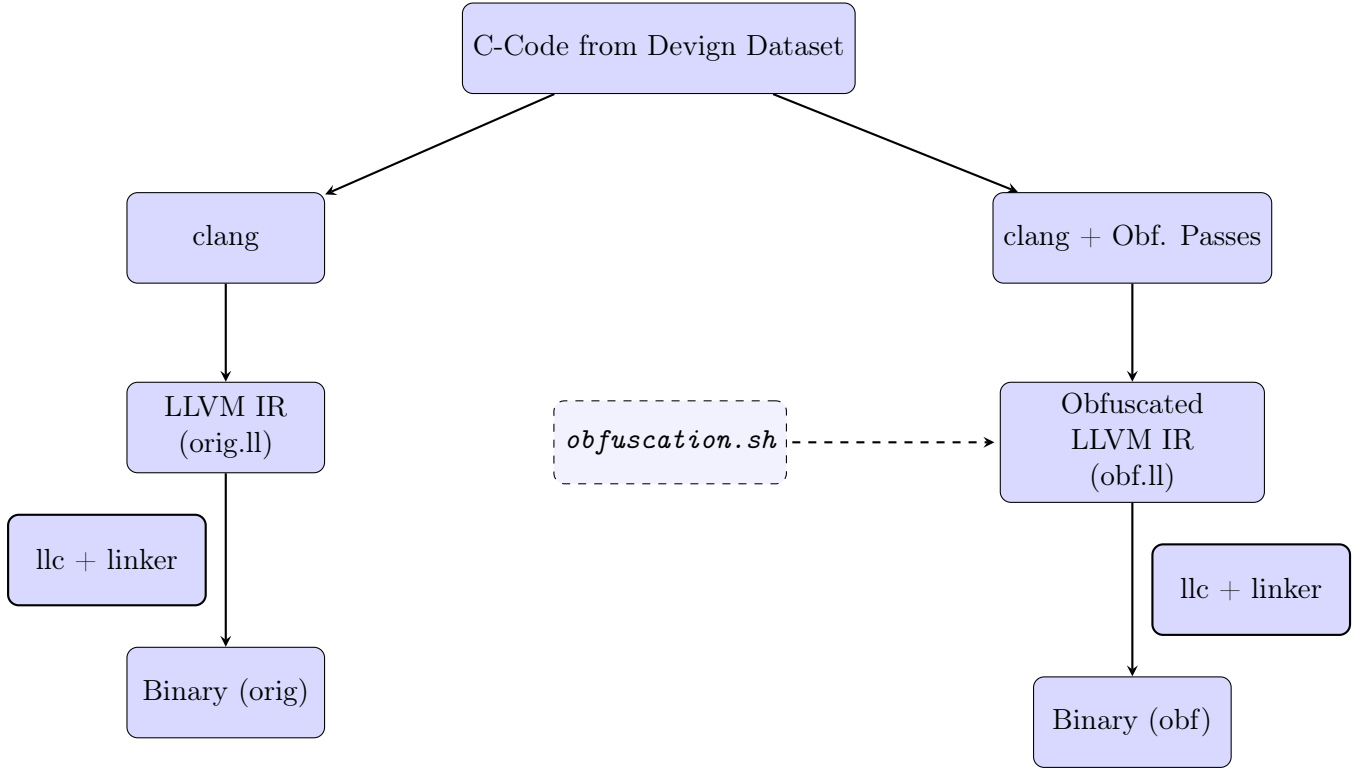
\begin{figure}[H]
\centering
\begin{tikzpicture}[
  node distance=1.3cm and 1.8cm,
  every node/.style={draw, rounded corners, fill=blue!15, align=center, font=\small, minimum height=1.2cm, minimum width=2.6cm},
  arrow/.style={thick, ->, >=stealth},
  dashedarrow/.style={thick, dashed, ->, >=stealth}
]

\node (cfile) {C-Code from Devign Dataset};

\node (clang) [below left=of cfile] {clang};
\node (clangobf) [below right=of cfile] {clang + Obf. Passes};

\node (llvmorig) [below=of clang] {LLVM IR\\(orig.ll)};
\node (llvmobf) [below=of clangobf, text width=3.2cm] {Obfuscated LLVM IR\\(obf.ll)};

\node (binorig) [below=of llvmorig, yshift = -1cm] {Binary (orig)};
\node (binobf) [below=of llvmobf, yshift = -1cm] {Binary (obf)};

\node (script) [left=2.8cm of llvmobf, draw, dashed, fill=blue!5, font=\itshape\small, minimum width=2.4cm, minimum height=1.1cm, yshift=0cm] {\texttt{obfuscation.sh}};

\draw[arrow] (cfile) -- (clang);
\draw[arrow] (cfile) -- (clangobf);
\draw[arrow] (clang) -- (llvmorig);
\draw[arrow] (clangobf) -- (llvmobf);
\draw[arrow] (llvmorig) -- (binorig) node[midway, left, font=\small, xshift = -0.25cm] {llc + linker};
\draw[arrow] (llvmobf) -- (binobf) node[midway, right, font=\small, xshift = 0.25cm] {llc + linker};
\draw[dashedarrow, shorten >=2pt, shorten <=2pt] (script.east) -- (llvmobf.west);

\end{tikzpicture}
\caption{Transformation pipeline from C source code to original and obfuscated LLVM IR using \texttt{obfuscation.sh}. The IR is compiled to binary using \texttt{llc} and a linker.}
\label{fig:obfuscation-pipeline}
\end{figure}

This chapter presents an empirical study that evaluates the robustness of three LLMs: ChatGPT-4o, Claude Sonnet 4, and Gemini Flash 2.5 in analyzing C functions. Our benchmarking pipeline automates code transformation and obfuscation via a custom script named \texttt{obfuscate.sh}. This script also copies the intermediate files and artifacts into designated folders to ensure clean and organized output. The process is illustrated in Fig. \ref{fig:obfuscation-pipeline}.

\vspace{1em}
\subsection{Dataset and Pre-processing}
\label{DatasetandPre-processing}

We use a balanced subset of the Devign dataset~\cite{zhou2019devign}, consisting of 40 functions in total: 20 labeled as vulnerable and 20 as safe. Before applying obfuscation, we manually modified each function to make it compilable using clang. This involved defining missing types, headers, and struct declarations. However, we ensured that the original logic of each function remained unchanged to preserve the ground truth vulnerability labels.

\vspace{1em}
\subsection{Obfuscation Pipeline}
\label{ObfuscationPipeline}

We employ the open-source \href{https://github.com/eshard/obfuscator-llvm}{Obfuscator-LLVM}~\cite{obfuscatorllvm2014} , a C++-based tool compatible with LLVM-14. This toolchain is freely available, actively maintained, and easily reproducible across different environments and it provides four configurable obfuscation techniques:

\begin{table}[H]
\centering
\renewcommand{\arraystretch}{1.8}
\setlength{\tabcolsep}{10pt}
\renewcommand{\baselinestretch}{1.5}

\resizebox{\textwidth}{!}{%
\begin{tabular}{|p{5cm}|p{4.5cm}|p{7cm}|}
\hline
\textbf{Technique} & \textbf{Description} & \textbf{C Code Example} \\
\hline

\textbf{Bogus Control Flow (BCF)} &
Inserts opaque conditional branches that are never executed, increasing control flow complexity. Often used to mislead static analysis. & \begin{flushleft}\ttfamily
int add(int a, int b) \{\\
\ \ if (a == b \&\& a != b) \{\\
\ \ \ \ printf("Fake path");\\
\ \ \}\\
\ \ return a + b;\\
\}
\end{flushleft} \\
\hline

\textbf{Control Flow Flattening} &
Replaces structured control flow with a dispatcher loop and switch-case. Makes decompilation and CFG recovery difficult. &
\begin{flushleft}\ttfamily
int compute(int x) \{\\
\ \ int state = 0;\\
\ \ while (1) \{\\
\ \ \ \ switch(state) \{\\
\ \ \ \ \ \ case 0: x += 1; state = 1; break;\\
\ \ \ \ \ \ case 1: x *= 2; return x;\\
\ \ \ \}\\
\ \ \}\\
\}
\end{flushleft} \\
\hline

\textbf{Instruction Substitution} &
Replaces simple instructions with semantically equivalent sequences to avoid detection. Alters opcode patterns. &
\begin{flushleft}\ttfamily
// Original:\\
x = y + z;\\[0.5em]
// Obfuscated:\\
x = y;\\
x += z;
\end{flushleft} \\
\hline

\textbf{Basic Block Splitting} &
Splits a basic block into smaller ones with jumps in between to break recognizable patterns. &
\begin{flushleft}\ttfamily
int multiply(int a, int b) \{\\
\ \ int result = a * b;\\
\ \ goto part2;\\
part2:\\
\ \ return result;\\
\}
\end{flushleft} \\
\hline

\end{tabular}
}
\caption{Obfuscation techniques that were used in our study}
\label{tab:obfuscation_techniques}
\end{table}

\textcolor{blue}{Focusing on these four transformations does not narrow the scope of our study. Similar techniques are well known in the research community and have also been observed in real-world malware, making them highly relevant for evaluating the robustness of LLMs against adversarial code transformations. Our selection was made to cover different layers of obfuscation: control flow (bogus control flow, flattening), semantics (instruction substitution), and syntax (basic block splitting). The generated obfuscated outputs and binary files were consistent across runs and formed the basis for our LLM analysis. As LLMs cannot directly process binary files, we further used the LLVM intermediate representation to enable experimentation.}

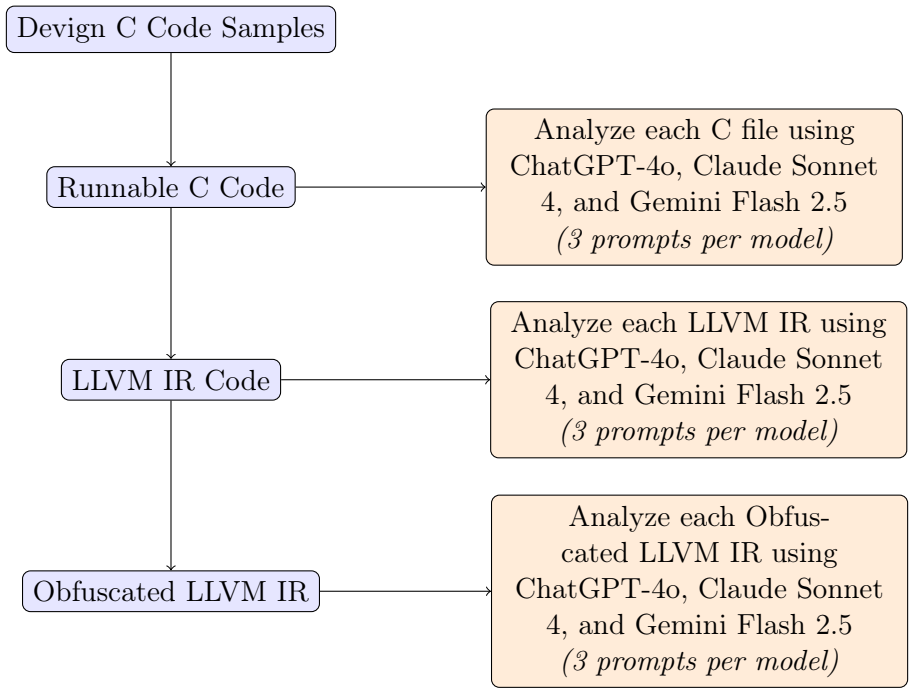
\begin{figure}[H]
\centering
\textbf{LLM Evaluation Pipeline for Vulnerability Detection}\\[1em] 
\begin{tikzpicture}[node distance=1.6cm and 3.5cm, font=\small, align=center]

\node (input) [draw, fill=blue!10, rounded corners=3pt] {Devign C Code Samples};
\node (c_code) [below=1.5cm of input, draw, fill=blue!10, rounded corners=3pt] {Runnable C Code};
\node (llvm) [below=2cm of c_code, draw, fill=blue!10, rounded corners=3pt] {LLVM IR Code};
\node (obf) [below=2.25cm of llvm, draw, fill=blue!10, rounded corners=3pt] {Obfuscated LLVM IR};

\node (c_eval) [right=of c_code, draw, xshift=-1cm, fill=orange!15, rounded corners=3pt, text width=5.2cm] 
{Analyze each C file using \\ ChatGPT-4o, Claude Sonnet 4, and Gemini Flash 2.5\\ \textit{(3 prompts per model)}};

\node (llvm_eval) [right=of llvm, draw, xshift=-0.75cm, fill=orange!15, rounded corners=3pt, text width=5.2cm] 
{Analyze each LLVM IR using \\ ChatGPT-4o, Claude Sonnet 4, and Gemini Flash 2.5\\ \textit{(3 prompts per model)}};

\node (obf_eval) [right=of obf, draw, xshift=-1.25cm, fill=orange!15, rounded corners=3pt, text width=5.2cm] 
{Analyze each Obfuscated LLVM IR using \\ ChatGPT-4o, Claude Sonnet 4, and Gemini Flash 2.5\\ \textit{(3 prompts per model)}};

\draw[->] (input) -- (c_code);
\draw[->] (c_code) -- (llvm);
\draw[->] (llvm) -- (obf);

\draw[->] (c_code) -- (c_eval);
\draw[->] (llvm) -- (llvm_eval);
\draw[->] (obf) -- (obf_eval);

\end{tikzpicture}
\caption{Evaluation pipeline for LLM-based vulnerability detection. The same samples are progressively transformed and analyzed using three different LLMs across multiple representations (C, LLVM IR, and obfuscated LLVM IR).}
\label{fig:llm_pipeline}
\end{figure}

These transformations were applied to the selected Devign functions. Obfuscation was fully automated using a custom script, which compiles the original functions, applies the selected passes via `opt`, and outputs transformed \texttt{.ll} and binary files. The process is illustrated in Fig. \ref{fig:llm_pipeline}. All scripts, data, configuration files, resulting obfuscated code, and evaluation results are made publicly available in our GitHub repository.\footnote{Available at \url{https://github.com/nikeboke/llm-llvm-adversarial}~\cite{llmllvmadversarial2025}}

\vspace{1em}
\subsection{Evaluation Setup}
\label{EvaluationSetup}

To evaluate the robustness of LLMs against compiler-level obfuscation, we tested three state-of-the-art models: ChatGPT-4o, Gemini Flash 2.5, and Claude Sonnet 4. Each function, both in its original and obfuscated form, was submitted to the models in a clean zero shot setting. Additionally, the corresponding .c source files were provided to the models prior to the LLVM IR representations in order to ensure that the models correctly understood the vulnerability status of the function before any compilation or transformation was applied. Due to the generative and non-deterministic nature of LLMs, we repeated each query three times for every function and used the majority prediction (`secure` or `insecure`) as the final classification. During experimentation, we observed that as chat sessions became longer, the models began to hallucinate due to memory leakage or context accumulation. To mitigate this effect and preserve the integrity of the zero-shot setting, each result was submitted in a separate, fresh chat session, and previous sessions were deleted. The prompt we used to query LLMs can be found in ~\ref{AppendixBLLMPromptTemplates}.  We recorded true positives (TP) and false negatives (FN) for both the original and obfuscated vulnerable samples.

\vspace{1em}
\subsection{Results and Observations}
\label{ResultsandObservations}

The results reveal a notable decline in vulnerability detection performance across all three models following obfuscation. For the original un-obfuscated functions, ChatGPT-4o achieved an accuracy of 0.800 with a precision of 0.772, recall of 0.850, specificity of 0.750, and F1-score of 0.810. Gemini Flash 2.5 followed with an accuracy of 0.675, and Claude Sonnet 4 at 0.650, both showing balanced recall and specificity. After obfuscation, ChatGPT-4o's accuracy dropped significantly to 0.525, and specificity fell to 0.250, indicating a much higher false positive rate. Gemini Flash 2.5 also showed reduced performance, with precision decreasing from 0.652 to 0.551 and specificity from 0.600 to 0.350. Claude Sonnet 4, while maintaining a strong recall of 0.900 after obfuscation, experienced a drop in specificity from 0.600 to 0.300. Further results can be found in Tab.~\ref{tab:llm_performance_comparison}. These findings underscore how compiler-level obfuscation impairs LLM-based vulnerability detection, particularly by reducing the models' ability to distinguish vulnerable code.

\begin{table}[H]
\centering
\renewcommand{\arraystretch}{1.3}
\setlength{\tabcolsep}{5pt}
\resizebox{\textwidth}{!}{%
\begin{tabular}{|l|c|c|c|c|c|c|}
\hline
\textbf{Model} & \textbf{Obfuscation} & \textbf{Accuracy} & \textbf{Precision} & \textbf{Recall (TPR)} & \textbf{Specificity (TNR)} & \textbf{F1-Score} \\
\hline
\multirow{2}{*}{ChatGPT-4o} 
& Before & 0.800 & 0.772 & 0.850 & 0.750 & 0.810 \\
& After  & 0.525 & 0.516 & 0.800 & 0.250 & 0.627 \\
\hline
\multirow{2}{*}{Gemini Flash 2.5} 
& Before & 0.675 & 0.652 & 0.750 & 0.600 & 0.698 \\
& After  & 0.575 & 0.551 & 0.800 & 0.350 & 0.653 \\
\hline
\multirow{2}{*}{Claude Sonnet 4} 
& Before & 0.650 & 0.636 & 0.700 & 0.600 & 0.666 \\
& After  & 0.600 & 0.560 & 0.900 & 0.300 & 0.692 \\
\hline
\end{tabular}%
}
\caption{Performance of LLMs before and after obfuscation.}
\label{tab:llm_performance_comparison}
\end{table}

Overall, our findings highlight a significant vulnerability in the current generation of LLM-based malware and vulnerability detectors. These models are effective on clean, human-readable code, but their reliability degrades substantially under compiler level adversarial transformations. This gap underscores the need for training LLMs on compiler level representations and for integrating more robust adversarial defense mechanisms in AI-driven security pipelines.

\subsection{Comparison Across LLMs}
\label{ComparisonAcrossLLMs}

Our comparative evaluation of ChatGPT-4o, Claude Sonnet 4, and Gemini Flash 2.5 reveals substantial differences in how each model responds to compiler-level obfuscation. While all three LLMs perform well on clean, unobfuscated functions, their robustness declines once LLVM-based transformations are applied.

Before obfuscation, ChatGPT-4o demonstrated the strongest overall performance with the highest accuracy of 0.800, precision of 0.772, recall of 0.850, specificity of 0.750, and F1-score of 0.810. Gemini Flash 2.5 followed with an accuracy of 0.675, showing balanced recall and precision, while Claude Sonnet 4 achieved slightly lower performance at 0.650 accuracy but with consistent scores across metrics.

After applying compiler-level obfuscation, performance dropped consistently across all models. ChatGPT-4o experienced a substantial decline in accuracy (0.800 to 0.525) and specificity (0.750 to 0.250), indicating a sharp increase in false positives. Claude Sonnet 4 achieved the highest recall post-obfuscation at 0.900, correctly identifying 18 out of 20 benign functions, but its specificity dropped to 0.300. Despite this trade-off, it ultimately achieved the highest F1-score after obfuscation (0.692). Gemini Flash 2.5 maintained a more balanced trade-off between recall (0.800) and specificity (0.350), offering a more moderate and consistent response to obfuscation.

These results underscore different model strengths: \textbf{ChatGPT-4o} was the most accurate and precise on clean code; \textbf{Claude Sonnet 4} prioritized vulnerability detection under obfuscation, achieving the highest recall and F1-score; and \textbf{Gemini Flash 2.5} balanced recall and specificity more evenly, making it a stable choice under adversarial conditions.

These results underscore that while LLMs retain some ability to detect obfuscated code patterns, their capacity to reliably identify vulnerabilities degrades significantly. Obfuscation induces a shift in model behavior toward falsely labeling vulnerable functions as benign, thereby increasing false negatives. This undermines the practical reliability of these models in real-world secure code bases, where undetected vulnerabilities pose serious risks.

In conclusion, future LLM-based security systems must incorporate adversarial robustness into their evaluation benchmarks. Training on obfuscated or intermediate representations, such as LLVM IR, and integrating obfuscation-aware pretraining may help counteract the adversarial effects of compiler-level transformations.

\section{Limitations}
\label{Limitations}

\textcolor{blue}{This study is subject to several limitations that should be considered when interpreting the results. First, the analysis relies exclusively on the three LLMs ChatGPT-4o, Gemini Flash 2.5, and Claude Sonnet 4. The restriction to these LLMs inherently limits the generalizability of our findings, as other LLMs may behave differently when confronted with the same tasks.}

\textcolor{blue}{Second, the generative nature of LLMs introduces variability in the answers they produce. Outputs can differ across runs even under identical input conditions, raising challenges for reproducibility and consistency. To mitigate this, we conducted three passes for each file; however, this repetition may not be sufficient to fully capture the variability of generative outputs. In addition, our analysis was based exclusively on zero-shot prompting, which, while providing an unbiased baseline, may not reflect the full range of model capabilities under different prompting strategies.}

\textcolor{blue}{Third, we focus only on \texttt{.ll}-files as part of the LLVM IR representation and not on the binary files (e.g., .exe). This choice was necessary because the evaluated LLMs are currently not able to directly process binary files and would first need to be fine-tuned for that purpose, which lies beyond the scope of this study. Furthermore, reverse engineering binary files into \texttt{.ll}-files is also not part of our experimental pipeline, as this step would exceed the scope of the work. While this ensures feasibility and consistency in the experimental setup, it also narrows the applicability of our results, as real-world usage often involves only the binary files}\footnote{Toor~\cite{toor2022ghidra} provides a solution to obtain \texttt{.ll}-files from binary files}.

\textcolor{blue}{Fourth, our evaluation is limited to static analysis only. While hybrid or dynamic analyses could provide deeper insights into runtime behavior and resistance to obfuscation, they were deliberately excluded from the scope of this work. Current large-language models are not yet well equipped to process runtime artifacts such as syscall traces, memory states, or dynamic control-flow graphs in a reliable and generalizable manner, particularly in zero-shot evaluation settings. As such, we focused on static code inspection and outline hybrid analysis as a promising avenue for future research.}

\textcolor{blue}{Fifth, we restricted the evaluation to the four obfuscation techniques that were implemented in our tool. Although this constraint ensures technical feasibility and consistency, it possibly reduces the breadth of the conclusions, as other obfuscation strategies were not considered.}

\textcolor{blue}{Sixth, we only use 40 C source files from the Devign dataset, which limits the scope of our analysis. However, limiting ourselves to a subset cannot cover the entire diversity and complexity of software vulnerabilities, which limits the generalizability of our results and may introduce dataset-specific biases.}

\section{Conclusion \& Outlook}
\label{ConclusionOutlook}

Our findings show that while LLMs such as ChatGPT-4o, Gemini Flash 2.5, and Claude Sonnet 4 can reliably detect vulnerabilities in clean C code, their effectiveness degrades significantly under compiler-level obfuscation. Basic syntactic and semantic obfuscation often fails to deceive these models, but transformations at the LLVM IR level consistently cause misclassifications in both vulnerable and safe functions. This shows that LLMs are still highly sensitive to changes in structural code, even when functionality is preserved.

These findings align with recent studies that reveal similar weaknesses. Prior work~\cite{secLLMHolmes} shows that even state-of-the-art models can yield incorrect answers in a significant number of cases due to small changes in code structure or naming. Further evidence~\cite{steenhoek2024error} demonstrates that LLMs struggle to reason about the code semantics essential for vulnerability identification, and that improvements in prompts, model size, or fine-tuning have not yet resolved these issues.

Given these limitations, we emphasize the need for LLMs that are specifically trained to understand and reason over low-level code representations, such as LLVM IR. Unlike general-purpose models, such specialized LLMs could be more resilient to compiler-level obfuscation and better equipped to detect embedded vulnerabilities. Developing such models represents a promising direction for future research toward more robust and trustworthy AI-driven security analysis.

Furthermore, compiler transformations can contribute to the adversarial hardening of LLM-based security tools. By systematically generating structurally diverse yet semantically equivalent code samples, researchers can create more comprehensive training and evaluation datasets. This approach can improve model generalization and reduce reliance on superficial code patterns. Recent work has shown that code-oriented language models are vulnerable to adversarially perturbed code generated via structural transformations, as demonstrated by Jha and Reddy~\cite{jha2022codeattack}. Complementing these findings, Bielik and Vechev proposed adversarial training and representation refinement techniques to improve the robustness of code models against semantically-preserving attacks~\cite{bielik2020adversarial}. Additionally, complementary methods such as self‑evaluation without fine-tuning have proven effective in reducing the success rate of adversarial inputs against both open-source and closed-source LLMs~\cite{brown2024selfeval}. These directions underscore the broader potential of compiler technologies in advancing software protection, code attribution, and the development of secure and reliable AI systems.

In addition to improving detection capabilities, compiler-based obfuscation techniques can also be explored as defensive mechanisms. For instance, intentional obfuscation during compilation can help protect proprietary software from reverse engineering and unauthorized access~\cite{collberg1997taxonomy}. Software watermarking represents another relevant application, where unique and resilient identifiers are embedded at the compiler level to trace software origin and detect unauthorized redistribution~\cite{collberg2009surreptitious}.

As part of our commitment to transparency and reproducibility, all artifacts used in our experiments, including code, scripts, prompt templates, and LLVM pass configurations, have been curated and will be made publicly available under an open license~\cite{llmllvmadversarial2025}. These resources are intended to support replication and enable the research community to build upon our findings.

\appendix
\section{Appendix A. Code Obfuscation Examples}
\label{CodeObfuscationExamples}

    \begin{table}[H]
    \centering
    \renewcommand{\arraystretch}{2.5} 
    \setlength{\tabcolsep}{16pt} 
    \renewcommand{\baselinestretch}{1.9} 
    \resizebox{1.1\textwidth}{!}{
    \begin{tabular}{|p{6cm}|p{9cm}|p{9cm}|} 
        \hline
        \textbf{\Large Technique} & \textbf{\Large Description} & \textbf{\Large C Code Example} \\
        \hline
        \textbf{\Large Variable Renaming} & 
        \parbox{9cm}{\vspace{10mm} \Large Changes variable and function names to meaningless identifiers to mislead static analysis tools. \vspace{10mm}} & 
        \parbox{9cm}{\vspace{10mm} \Large \texttt{
        int t9s\_y1(int x1, int x2) \{ \\
        \ \ return x1 + x2; \}} \vspace{10mm}} \\
        \hline
        \textbf{\Large Dead Code Insertion} & 
        \parbox{9cm}{\vspace{10mm} \Large Adds unnecessary code that does not affect functionality but confuses analysis tools. \vspace{10mm}} & 
        \parbox{9cm}{\vspace{10mm} \Large \texttt{
        int calculate(int a, int b) \{ \\
        \ \ int x = 10; // unused variable \\
        \ \ int y = x * 5; // redundant calculations \\
        \ \ if (y > 50) printf("Good for nothing."); \\
        \ \ return a + b; \}} \vspace{10mm}} \\
        \hline
        \textbf{\Large Formatting Obfuscation} & 
        \parbox{9cm}{\vspace{10mm} \Large Alters code formatting (spacing, indentation) while keeping logic unchanged. \vspace{10mm}} & 
        \parbox{9cm}{\vspace{10mm} \Large \texttt{
        int calculate(int a,int b)\{return a+b;\}} \vspace{10mm}} \\
        \hline
    \end{tabular}
    }
    \caption{\normalsize Examples of Syntactic Obfuscation Techniques in C}
    \label{tab:syntactic_obfuscation}
\end{table}

\begin{table}[H] 
    \centering
    \renewcommand{\arraystretch}{1.8} 
    \setlength{\tabcolsep}{10pt} 
    \renewcommand{\baselinestretch}{1.5} 

    \resizebox{\textwidth}{!}{
    \begin{tabular}{|p{5cm}|p{6.5cm}|p{5cm}|} 
        \hline
        \textbf{\small Technique} & \textbf{\small Description} & \textbf{\small C Code Example} \\
        \hline
        \textbf{\small Control Flow Distortion} & 
        \parbox{5.5cm}{\vspace{5mm} \small Alters the program’s execution flow, making it harder to analyze and understand the logic. \vspace{5mm}} & 
        \parbox{5.5cm}{\vspace{5mm} \small \texttt{
        int calculate(int a, int b) \{ \\
        \ \ int state = 0, result; \\
        \ \ while (1) \{ \\
        \ \ \ \ switch (state) \{ \\
        \ \ \ \ \ \ case 0: state = 1; break; \\
        \ \ \ \ \ \ case 1: result = a + b; state = 2; break; \\
        \ \ \ \ \ \ case 2: return result; \\
        \ \ \ \ \} \\
        \ \ \} \}} \vspace{5mm}} \\
        \hline
        \textbf{\small Opaque Predicate} & 
        \parbox{5.5cm}{\vspace{5mm} \small Introduces conditional statements that always evaluate to the same value to confuse analysis tools. \vspace{5mm}} & 
        \parbox{5.5cm}{\vspace{5mm} \small \texttt{
        int calculate(int a, int b) \{ \\
        \ \ int x = 1337; \\
        \ \ if ((x * x) \% 1337 == 0) // always true \\
        \ \ \ \ return a + b; \\
        \ \ else \\
        \ \ \ \ return a - b; \}} \vspace{5mm}} \\
        \hline
        \textbf{\small Instruction Substitution} & 
        \parbox{5.5cm}{\vspace{5mm} \small Replaces simple operations with more complex equivalents to obscure intent. \vspace{5mm}} & 
        \parbox{5.5cm}{\vspace{5mm} \small \texttt{
        int calculate(int a, int b) \{ \\
        \ \ int result = 0; \\
        \ \ for (int i = 0; i < abs(b); i++) \\
        \ \ \ \ result += a; \\
        \ \ if (b < 0) result = -result; \\
        \ \ return result; \}} \vspace{5mm}} \\
        \hline
    \end{tabular}
    }

    \caption{\small Examples of Semantic Obfuscation Techniques in C}
    \label{tab:semantic_obfuscation}
\end{table}

\noindent 
\begin{table}[H] 
    \centering
    \renewcommand{\arraystretch}{1.8}
    \setlength{\tabcolsep}{10pt} 
    \renewcommand{\baselinestretch}{1.5}

    \resizebox{\textwidth}{!}{
    \begin{tabular}{|p{5cm}|p{6.5cm}|p{5.5cm}|} 
        \hline
        \textbf{\small Technique} & \textbf{\small Description} & \textbf{\small C Code Example} \\
        \hline
        \textbf{\small Packing/Encryption} & 
        \parbox{4.5cm}{\vspace{3mm} \small Encrypts or compresses code, making it unreadable until runtime, when it is dynamically unpacked. \vspace{5mm}} & 
        \parbox{4.5cm}{\vspace{5mm} \small \texttt{
        void encrypted\_payload() \{ \\
        \ \ char payload[] = "ajdjfjfjfj"; \\
        \ \ int key = 847584; \\
        \ \ for (int i = 0; i < strlen(payload); i++) \\
        \ \ \ \ payload[i] \textasciicircum= key; \\
        \ \ printf("Decrypted: \%s", payload); \}} \vspace{5mm}} \\
        \hline
        \textbf{\small Polymorphic Behavior} & 
        \parbox{4.5cm}{\vspace{5mm} \small Code dynamically mutates its structure while maintaining the same functionality, making signature-based detection difficult. \vspace{5mm}} & 
        \parbox{4.5cm}{\vspace{5mm} \small \texttt{
        void poly\_function(int a, int b) \{ \\
        \ \ int (*func\_ptr)(int, int); \\
        \ \ if ((rand() \% 2) == 0) func\_ptr = \&calculate; \\
        \ \ else func\_ptr = \&alternate\_calculate; \\
        \ \ int result = func\_ptr(a, b); \\
        \ \ printf("Result: \%d", result); \}} \vspace{5mm}} \\
        \hline
        \textbf{\small Metamorphic Code} & 
        \parbox{4.5cm}{\vspace{5mm} \small Rewrites itself during execution, ensuring that no two versions of the code are identical, even though behavior remains unchanged. \vspace{5mm}} & 
        \parbox{7.5cm}{\vspace{5mm} \small \texttt{
        void metamorphic\_function()\{ \\
        \ \ char *variants[] = \{"Char1", "Char2"\}; \\
       \ \ int choice = rand() \% 2; \\  
        \ \ printf("Executing: \%s", variants[choice]); \}} \vspace{5mm}} \\
        \hline
    \end{tabular}
    }

    \caption{\small Examples of Behavioral Obfuscation Techniques in C}
    \label{tab:behavioral_obfuscation}
\end{table}

\clearpage

\renewcommand{\thetable}{B.\arabic{table}}  
\setcounter{table}{0}  

\section{Appendix B. LLM Prompt Templates}
\label{AppendixBLLMPromptTemplates}

\begin{table}[h!]
\centering
\begin{tabular}{|p{3cm}|p{10cm}|}
\hline
\textbf{Prompt Type} & \textbf{Prompt Text} \\
\hline
C-Code &
Analyze the provided C source code for potential security vulnerabilities or signs of malicious behavior. Using exactly one of the following formats:
\begin{itemize}
  \item If the file is secure: Yes, the code is secure.
  \item If the file is insecure: No, the code is insecure because [reason].
\end{itemize}
Answer briefly.

Code:
\\
\hline
LLVM-orig-code &
Analyze the provided LLVM source code for potential security vulnerabilities or signs of malicious behavior. Using exactly one of the following formats:
\begin{itemize}
  \item If the file is secure: Yes, the code is secure.
  \item If the file is insecure: No, the code is insecure because [reason].
\end{itemize}
Answer briefly.

Code:
\\
\hline
LLVM-obf-code &
Analyze the provided obfuscated LLVM source code for potential security vulnerabilities or signs of malicious behavior. Using exactly one of the following formats:
\begin{itemize}
  \item If the file is secure: Yes, the code is secure.
  \item If the file is insecure: No, the code is insecure because [reason].
\end{itemize}
Answer briefly.

Code:
\\
\hline
\end{tabular}
\caption{Prompts used for each code type evaluated by the LLMs}
\label{tab:prompts}
\end{table}

\section*{Declaration of Generative AI and AI-Assisted Technologies in the Writing Process}
\label{DeclarationofGenerativeAIandAI-AssistedTechnologiesintheWriting Process}

During the preparation of this work the authors used ChatGPT-4o, Gemini Flash 2.5 and Claude Sonnet 4 to assist with code analysis and grammar refinement. After using this tool/service, the authors reviewed and edited the content as needed and take full responsibility for the content of the publication.

\clearpage

\bibliographystyle{elsarticle-num}
\bibliography{references}

\begin{thebibliography}{10}
\expandafter\ifx\csname url\endcsname\relax
  \def\url#1{\texttt{#1}}\fi
\expandafter\ifx\csname urlprefix\endcsname\relax\def\urlprefix{URL }\fi
\expandafter\ifx\csname href\endcsname\relax
  \def\href#1#2{#2} \def\path#1{#1}\fi

\bibitem{lyu2024codeexecutors}
C.~Lyu, L.~Yan, R.~Xing, W.~Li, Y.~Samih, T.~Ji, L.~Wang,
  \href{https://arxiv.org/abs/2410.06667}{Large language models as code
  executors: An exploratory study}, arXiv preprint arXiv:2410.06667 (2024).
\newline\urlprefix\url{https://arxiv.org/abs/2410.06667}

\bibitem{llvm}
{LLVM Project}, \href{https://llvm.org/docs/LangRef.html}{LLVM Language
  Reference Manual} (2025).
\newline\urlprefix\url{https://llvm.org/docs/LangRef.html}

\bibitem{toor2022ghidra}
T.~S. Toor,
  \href{https://uwspace.uwaterloo.ca/bitstream/handle/10012/17976/Toor_Tejvinder.pdf}{Decompilation
  of binaries into llvm ir for automated analysis}, Master's thesis, University
  of Waterloo (2022).
\newline\urlprefix\url{https://uwspace.uwaterloo.ca/bitstream/handle/10012/17976/Toor_Tejvinder.pdf}

\bibitem{ghidra}
{National Security Agency}, {Ghidra Software Reverse Engineering Framework},
  \url{https://ghidra-sre.org/} (2019).

\bibitem{openai_chatgpt4o}
OpenAI, Chatgpt-4o model by openai, \url{https://chat.openai.com}, accessed
  March 2025 (2024).

\bibitem{gemini2025}
G.~DeepMind, Gemini 2.5, \url{https://deepmind.google/technologies/gemini/},
  accessed: July 2025 (2024).

\bibitem{claude2025}
Anthropic, Introducig claude 4, \url{https://www.anthropic.com/news/claude-4},
  accessed: July 2025 (2024).

\bibitem{obfuscatorllvm2014}
{Obfuscator-LLVM Team}, {Obfuscator-LLVM},
  \url{https://github.com/eshard/obfuscator-llvm}, accessed: 2025-02-24 (2014).

\bibitem{revay2023cffvb}
G.~Revay,
  \href{https://www.virusbulletin.com/uploads/pdf/conference/vb2023/papers/Dont-flatten-yourself-restoring-malware-with-Control-Flow-Flattening-obfuscation.pdf}{Don't
  flatten yourself: restoring malware with control-flow flattening
  obfuscation}, in: Virus Bulletin Conference (VB2023), Virus Bulletin, London,
  United Kingdom, 2023.
\newline\urlprefix\url{https://www.virusbulletin.com/uploads/pdf/conference/vb2023/papers/Dont-flatten-yourself-restoring-malware-with-Control-Flow-Flattening-obfuscation.pdf}

\bibitem{zhou2019devign}
Y.~Zhou, S.~Liu, J.~Siow, X.~Du, Y.~Liu,
  \href{https://arxiv.org/abs/1909.03496}{Devign: Effective vulnerability
  identification by learning comprehensive program semantics via graph neural
  networks}, in: Advances in Neural Information Processing Systems (NeurIPS),
  Vol.~32, 2019.
\newline\urlprefix\url{https://arxiv.org/abs/1909.03496}

\bibitem{feature}
M.~Anderson, M.~Cafarella, Input selection for fast feature engineering, 2016,
  pp. 577--588.
\newblock \href {https://doi.org/10.1109/ICDE.2016.7498272}
  {\path{doi:10.1109/ICDE.2016.7498272}}.

\bibitem{raff2017malware}
E.~Raff, J.~Barker, J.~Sylvester, R.~Brandon, B.~Catanzaro, Malware detection
  by eating a whole exe, arXiv preprint arXiv:1710.09435 (2017).

\bibitem{singh2024malicious}
R.~Singh, P.~Gupta, A.~Sharma, Malicious code detection using llm, IEEE
  Transactions on Dependable and Secure Computing 21~(1) (2024) 85--96.
\newblock \href {https://doi.org/10.1109/TDSC.2024.10670668}
  {\path{doi:10.1109/TDSC.2024.10670668}}.

\bibitem{brown2020language}
T.~B. Brown, B.~Mann, N.~Ryder, M.~Subbiah, J.~Kaplan, P.~Dhariwal,
  A.~Neelakantan, P.~Shyam, G.~Sastry, A.~Askell, S.~Agarwal, N.~Herbert-Voss,
  G.~Krueger, T.~Henighan, R.~Child, A.~Ramesh, D.~M. Ziegler, J.~Wu,
  C.~Winter, C.~Hesse, M.~Chen, E.~Sigler, M.~Litwin, S.~Gray, B.~Chess,
  J.~Clark, C.~Berner, S.~McCandlish, A.~Radford, I.~Sutskever, D.~Amodei,
  Language models are few-shot learners, Advances in Neural Information
  Processing Systems 33 (2020).

\bibitem{devlin2018bert}
J.~Devlin, M.-W. Chang, K.~Lee, K.~Toutanova, Bert: Pre-training of deep
  bidirectional transformers for language understanding, arXiv preprint
  arXiv:1810.04805 (2018).
\newblock \href {https://doi.org/10.48550/arXiv.1810.04805}
  {\path{doi:10.48550/arXiv.1810.04805}}.

\bibitem{Feng:codebert}
Z.~Feng, D.~Guo, D.~Tang, N.~Duan, X.~Feng, M.~Gong, L.~Shou, T.~Liu, D.~Jiang,
  M.~Zhou, Codebert: A pre-trained model for programming and natural languages,
  2020, pp. 1536--1547.
\newblock \href {https://doi.org/10.18653/v1/2020.findings-emnlp.139}
  {\path{doi:10.18653/v1/2020.findings-emnlp.139}}.

\bibitem{lu2021codexglue}
S.~Lu, D.~Guo, S.~Ren, J.~Huang, A.~Svyatkovskiy, A.~Blanco, C.~Clement,
  D.~Drain, D.~Jiang, D.~Tang, et~al., Codexglue: A machine learning benchmark
  dataset for code understanding and generation, arXiv preprint
  arXiv:2102.04664 (2021).

\bibitem{li2024automated}
C.~Li, J.~Zhang, L.~Wang, X.~Xu, Automated vulnerability detection using deep
  learning techniques, arXiv preprint arXiv:2410.21968 (2024).

\bibitem{Rahali2021MalBERT}
A.~Rahali, M.~A. Akhloufi, \href{https://arxiv.org/abs/2103.03806}{Malbert:
  Using transformers for cybersecurity and malicious software detection}, arXiv
  preprint arXiv:2103.03806 (2021).
\newline\urlprefix\url{https://arxiv.org/abs/2103.03806}

\bibitem{Allix2016Androzoo}
K.~Allix, T.~F. Bissyandé, J.~Klein, Y.~L. Traon, Androzoo: Collecting
  millions of android apps for the research community, in: Proceedings of the
  13th International Conference on Mining Software Repositories, MSR '16, ACM,
  New York, NY, USA, 2016, pp. 468--471.
\newblock \href {https://doi.org/10.1145/2901739.2903508}
  {\path{doi:10.1145/2901739.2903508}}.

\bibitem{9785789}
D.~Demirci, N.~şahin, M.~şirlancis, C.~Acartürk, Static malware detection
  using stacked bilstm and gpt-2, IEEE Access 10 (2022) 58488--58502.
\newblock \href {https://doi.org/10.1109/ACCESS.2022.3179384}
  {\path{doi:10.1109/ACCESS.2022.3179384}}.

\bibitem{AkinsowonJiang2024}
T.~Akinsowon, H.~Jiang,
  \href{https://ihsonline.org/Portals/0/Tech%20Papers/2024_Papers/Akinsowon-Jiang_LeveragingLargeLanguageModels.pdf?ver=2RlcpRiD4bUoBIcb9SnOsg%3D%3D}{Leveraging
  large language models for behavior-based malware detection using deep
  learning}, IHS Online Technical PapersAccessed: 01.02.2025 (2024).
\newline\urlprefix\url{https://ihsonline.org/Portals/0/Tech%20Papers/2024_Papers/Akinsowon-Jiang_LeveragingLargeLanguageModels.pdf?ver=2RlcpRiD4bUoBIcb9SnOsg%3D%3D}

\bibitem{li2022multiview}
W.~Li, J.~Zhang, C.~Huang, M.~Wang,
  \href{https://arxiv.org/abs/2210.15429}{Multi-view representation learning
  from malware to defend against adversarial variants}, arXiv preprint
  arXiv:2210.15429 (2022).
\newline\urlprefix\url{https://arxiv.org/abs/2210.15429}

\bibitem{lattner2004llvm}
C.~Lattner, V.~Adve, \href{https://llvm.org/pubs/2004-01-30-CGO-LLVM.pdf}{Llvm:
  A compilation framework for lifelong program analysis and transformation},
  in: Proceedings of the 2004 International Symposium on Code Generation and
  Optimization (CGO'04), IEEE Computer Society, 2004, pp. 75--86.
\newblock \href {https://doi.org/10.1109/CGO.2004.1281665}
  {\path{doi:10.1109/CGO.2004.1281665}}.
\newline\urlprefix\url{https://llvm.org/pubs/2004-01-30-CGO-LLVM.pdf}

\bibitem{sharif2008impeding}
M.~Sharif, A.~Lanzi, J.~Giffin, W.~Lee,
  \href{https://llvm.org/pubs/2008-02-ImpedingMalwareAnalysis.pdf}{Impeding
  malware analysis using conditional code obfuscation}, in: Network and
  Distributed System Security Symposium (NDSS), 2008.
\newline\urlprefix\url{https://llvm.org/pubs/2008-02-ImpedingMalwareAnalysis.pdf}

\bibitem{idapro}
Hex-Rays, Ida pro disassembler, \url{https://www.hex-rays.com/products/ida/}
  (2023).

\bibitem{junod2015obfuscator}
P.~Junod, J.~Rinaldini, J.~Wehrli, J.~Michielin, Obfuscator-llvm: Software
  protection for the masses, in: Proceedings of the IEEE/ACM 1st International
  Workshop on Software Protection (SPRO'15), IEEE, 2015, pp. 3--9.

\bibitem{brezinski2023metamorphic}
K.~Brezinski,
  \href{https://onlinelibrary.wiley.com/doi/10.1155/2023/8227751}{Metamorphic
  malware and obfuscation: A survey of obfuscation techniques used by mutation
  engines}, ISRN Computer Science 2023 (2023) 1--19.
\newblock \href {https://doi.org/10.1155/2023/8227751}
  {\path{doi:10.1155/2023/8227751}}.
\newline\urlprefix\url{https://onlinelibrary.wiley.com/doi/10.1155/2023/8227751}

\bibitem{zhang2023khaos}
P.~Zhang, C.~Wu, M.~Peng, K.~Zeng, D.~Yu, Y.~Lai, Y.~Kang, W.~Wang, Z.~Wang,
  \href{https://arxiv.org/abs/2301.11586}{Khaos: The impact of inter-procedural
  code obfuscation on binary diffing techniques}, arXiv preprint
  arXiv:2301.11586Submitted 27 Jan 2023 (2023).
\newblock \href {https://doi.org/10.48550/arXiv.2301.11586}
  {\path{doi:10.48550/arXiv.2301.11586}}.
\newline\urlprefix\url{https://arxiv.org/abs/2301.11586}

\bibitem{vmware2019apt10}
{VMware Threat Analysis Unit}, Defeating compiler-level obfuscations used in
  apt10 malware,
  \url{https://blogs.vmware.com/security/2019/02/defeating-compiler-level-obfuscations-used-in-apt10-malware.html},
  accessed: 2025-09-03 (2019).

\bibitem{llmllvmadversarial2025}
{Böke, Ekin and Torka, Simon}, {LLM-LLVM-Adversarial},
  \url{https://github.com/nikeboke/llm-llvm-adversarial}, accessed: 2025-09-13
  (2025).

\bibitem{secLLMHolmes}
Y.~Li, C.~Li, Y.~Li, K.~Chen, Y.~Zhang, Y.~Wang, S.~Zhu, X.~Wang,
  \href{https://arxiv.org/abs/2312.12575}{Secllmholmes: Probing llms for
  security vulnerability detection and reasoning}, arXiv preprint
  arXiv:2312.12575 (2023).
\newline\urlprefix\url{https://arxiv.org/abs/2312.12575}

\bibitem{steenhoek2024error}
W.~Steenhoek, Y.~Siwakoti, M.~Yuan, D.~Marinov,
  \href{https://arxiv.org/abs/2403.17218}{To err is machine: Vulnerability
  detection challenges llm reasoning}, arXiv preprint arXiv:2403.17218 (2024).
\newline\urlprefix\url{https://arxiv.org/abs/2403.17218}

\bibitem{jha2022codeattack}
A.~Jha, C.~K. Reddy, Codeattack: Code-based adversarial attacks for pre-trained
  programming language models, arXiv preprint arXiv:2206.00052 (2022).

\bibitem{bielik2020adversarial}
P.~Bielik, M.~Vechev, Adversarial robustness for code, arXiv preprint
  arXiv:2002.04694 (2020).

\bibitem{brown2024selfeval}
H.~Brown, L.~Lin, K.~Kawaguchi, M.~Shieh,
  \href{https://arxiv.org/abs/2407.03234}{Self-evaluation as a defense against
  adversarial attacks on llms}, arXiv preprint arXiv:2407.03234 (2024).
\newline\urlprefix\url{https://arxiv.org/abs/2407.03234}

\bibitem{collberg1997taxonomy}
C.~Collberg, C.~Thomborson, D.~Low, A taxonomy of obfuscating transformations,
  Tech. Rep. 148, University of Auckland (1997).

\bibitem{collberg2009surreptitious}
C.~Collberg, J.~Nagra, Surreptitious software: obfuscation, watermarking, and
  tamperproofing for software protection, Addison-Wesley, 2009.

\end{thebibliography}

\end{document}